\documentclass[14pt,a4paper,useAMS,usenatbib]{emulateapj}
\usepackage{epsfig}
\usepackage{amsmath}
\usepackage{natbib}

\begin{document}

\title{Formation of different isotopomers of chloronium in the interstellar medium}

\author{Liton Majumdar\altaffilmark{1}, Ankan Das\altaffilmark{1}, Sandip K. Chakrabarti\altaffilmark{2,1}}
\affil{$^1$Indian Centre for Space Physics, Chalantika 43, Garia Station Rd., Kolkata, 700084, India}
\affil{$^2$S. N. Bose National Centre for Basic Sciences, Salt Lake, Kolkata, 700098, India}

\email{$^1$ankan.das@gmail.com}

\shorttitle{Formation of Chloronium ion in ISM}
\shortauthors{Majumdar et al.}

\begin{abstract}
Main focus of this paper is to explore the possibility of findings two deuterated
isotopomers of H$_2$Cl$^+$ (Chloronium) in and around the Interstellar Medium (ISM).
Presence of Chloronium ion has recently been confirmed by
Herschel Space Observatory$'$s Heterodyne Instrument     
for the Far-Infrared \citep{neuf12}. It observed para-chloronium towards six sources in
the Galaxy. Existence of its deuterated isotopomers (HDCl$^+$ \& D$_2$Cl$^+$) are till date not
discussed in the literature. 
We find that these deuterated gas phase ions could be destroyed by 
various ion-molecular reactions, dissociative recombination (DR) and by cosmic rays (CR). 
We compute all the Ion-molecular (polar) reaction rates by using the parameterized trajectory theory
and the Ion-molecular (non-polar) reaction rates by using the Langevin theory.
For DR and CR induced reactions, we adopt two well behaved rate formulas. We also include 
these rate coefficients into our large gas-grain chemical network to study the
chemical evolution of these species around the outer edge of the cold dense cloud.
In order to study spectral properties of chloronium ion and its two deuterated
isotopomers, we have carried out quantum chemical simulations.
We calculated ground state properties of these species by employing second order
Moller-Plesset perturbation theory (MP2) along with quadruple-zeta correlation consistent (aug-cc-pVQZ) 
basis set.
Infrared and electronic absorption spectra of these species are calculated by using
the same level of theory. MP2/aug-cc-pVQZ level of theory is used to report the different 
spectroscopic constants of these gas phase species. These spectroscopic constants are essential to predict
rotational transitions of these species. Our predicted column densities of D$_2$Cl$^{+}$, HDCl$^{+}$ 
along with the spectral information may enable their future identification around outer edges of cold 
dark clouds.
\end{abstract}

\section{Introduction}
Cologne Database for Molecular Spectroscopy (CDMS) catalog \citep{mull01,mull05} 
keeps records on discovered molecules in interstellar medium (ISM). According this catalog, 
more than 170 molecules have been detected so far in the ISM. Among them, there are 
some simple halides like, such as, HF, CF, AlF, HCl, HCl$^+$, NaCl, KCl, AlCl, MgCl etc.
Several works have been done in the past to model the 
chemistry of the chlorine-bearing molecules in both the diffuse and dense molecular 
clouds \citep{jura74,dalg74,van86,schi95,fede95,amin96,neuf09}.
Various halogen elements such as fluorine and chlorine having solar abundances 
($3.6\times 10^{-8}$ and $3.2\times 10^{-7}$ respectively relative to total 
hydrogen nuclei \citep{aspl09}) play an important role towards the formation of various hydrides 
in the ISM. Dissociation energy of most of the hydrides are 
less than that of the hydrogen molecule, only exceptions are the diatomic hydrides, such as, HF 
and diatomic hydride cations, such as, HCl$^+$ \citep{lis10}. 
Around an ISM, most of the halogen elements are biased to form hydrides \citep{neuf10}.
Huge abundances of HF and H$_2$Cl$^+$ are observed around the diffuse molecular 
clouds \citep{neuf10,sonn10,lis10}. 

The ionization potential of chlorine is slightly less than that of the hydrogen and 
singly-ionized chlorine (Cl$^+$) can react exothermically with H$_2$ to form HCl$^+$. 
This HCl$^+$ further interacts with H$_2$ to form H$_2$Cl$^+$. 
These features are already known by theory of thermo chemistry of these species. 
Since Choloronium ion does not react with hydrogen molecule, dissociative recombination
appears to be its main destruction route. Chloronium ion could also be destroyed by reacting with CO
molecule. Main product of these destruction routes is hydrogen chloride (HCl). 
Theoretical modeling by considering the chlorine bearing molecules predicts that abundances of the
Chloronium ions are significantly higher and could be observed. But surprisingly,
before the launching of Herschel, among the Cl bearing molecules, only H$^{35}$Cl and H$^{37}$Cl were 
detected \citep{blak85,zmui95,schi95,sale96}. Chemical modeling suggests that Chloronium ions 
could be very abundant around the diffuse interstellar medium. The detection 
of H$_2$Cl$^+$ was first reported towards NGC 6334I and Sgr B2(S) using the HIFI instrument 
\citep{degr10} aboard the Herschel Space Observatory \citep{pilb10}. 
A follow up study by \citet{neuf12} also detected H$_2$Cl$^+$ absorption towards Sgr A, W31C
and detected Chloronium emission from two sources in the Orion Molecular Cloud 1 
(the Orion Bar photodissociation region and Orion South condensation) and 
the young massive star AFGL 2591.

Despite these overwhelmingly significant observational evidences, till date, no
deuterated forms of  H$_2$Cl$^+$ have been observed in the ISM. This motivates us to
model the formation/destruction of different forms of deuterated H$_2$Cl$^+$ in the ISM. 
Importance of interstellar grains in producing simpler molecules has been widely described by
several authors such as \citet{stan02,hase92,chak06a,chak06b,das08a,das08b,das10,das11}. 
These studies indicate that dusts could play a crucial role for deciding chemical compositions 
around any molecular cloud. In order to understand complete picture of how molecules are 
formed in or around an ISM, we used our large gas-grain chemical model 
\citep{das13a,maju12,maju13a,maju13b}.

\citet{maju13a,maju13b} \& \citet{das13b} performed quantum chemical simulations to find out various 
chemical properties of some interstellar species. In the current paper, we carry out 
a similar types of quantum chemical calculations to find out the spectral information 
(infrared, electronic and rotational) of different deuterated istotopomers of Chloronium ion. 
We have also provided detailed information about rotational transitions of these species
(in the format of JPL catalog). Such a study would be extremely helpful for the identification
of these species around ISMs.

The plan of this paper is the following: In Section 2, the models and the
computational details are presented. Implications of the results are discussed in
Section 3. Finally, in Section 4, we draw our conclusions. In Appendix A, we 
tabulate relevant parameters for rotational spectroscopy of one of the isotopomer of 
H$_2$Cl$^+$ in the format of JPL catalog.  

\section{Computational details}

\subsection{Quantum chemical simulations and derived spectral parameters}

We study spectral properties of interstellar chloronium by first optimizing the 
geometry of chloronium using second order Moller-Plesset perturbation theory (MP2) along with
aug-cc-PVQZ basis set. MP2 theory is a special case of more general many-body perturbation 
theory  \citep{puzz10}. For calculations where only valence electrons are correlated 
(i.e., frozen core calculation), standard cc-pVXZ sets with X= D, T, Q, 5 and 6 are recommended 
\citep{puzz10}.  Because of this, we use here aug-cc-pVQZ basis set along with
MP2 method. Vibrational frequencies of H$_2$Cl$^+$ and its two deuterated isotopomers; D$_2$Cl$^+$ and 
HDCl$^+$ are computed by determining second derivative of energy with respect to Cartesian 
nuclear coordinates and then transforming into mass-weighted coordinates. This transformation 
is valid only at a stationary point. 

Vibrational frequencies of these species in the ice phase are also obtained by using 
the same level of theory. Vibrational spectra of any chemical species are significantly affected by the
type of solvent used in the simulation. In our modeling purpose, here we have used
Polarizable Continuum Model (PCM) with integral equation formalism variant (IEFPCM) 
as a default Self-consistent Reaction Field (SCRF) method \citep{toma02,toma05,toma99,pasc94}. SCRF 
method in Gaussian 09W program is used to perform calculations in presence of a solvent by 
placing the solute in a cavity within a solvent reaction field.
Following \citet{das13b}, here too we have considered simple as well as mixed ice. By 
simple ice, we mean ice made from water molecule only and by mixed ice, we mean the ice made from
water, methanol and carbon di-oxide. Observations around cold dense region of 
molecular cloud reveal that $\sim 90$\% of interstellar grain mantle could be  
covered by H$_2$O, CH$_3$OH and CO$_2$ \citep{kean01}. We have also carried 
out quantum chemical simulation (time dependent density functional theory) to 
find out electronic absorption spectrum of these species in the gas phase as 
well as simple ice and mixed ice phase using the IEFPCM model.

An important  key for the successful identification of different 
forms of chloronium in the interstellar space depends on the 
availability of accurate predictions of spectroscopic constants
(rotational and distortional constants). Our computed rotational spectral information for
interstellar chloronium is based on equilibrium geometry obtained at MP2/aug-cc-pVQZ level, 
i.e., with the consideration of core correlation and vibrational corrections to the 
rotational constants. Centrifugal distortion constants are computed 
from harmonic and anharmonic force fields obtained at 
MP2/aug-cc-pVQZ levels of theory. In addition, computed components of dipole moment were 
used to predict relative intensities of rotational transitions. These spectroscopic constants 
are essential to predict spectrum of different forms of H$_2$Cl$^+$ and this can be done by following 
techniques described in \citet{das13b}. It used Picket's `SPCAT' program \citep{pick91}
for this purpose. Two main files, namely, `file.var' and `file.int' are required
for the SPCAT program. Specified format of these two files were explained in  
detail in \citet{pick91}. Information regarding rotational constants, 
quadrupole coupling constants and distortional constants are given in `.var' file. Contents of 
`.var' file could directly be generated from Gaussian 09 program. 
`.int' file is the 
intensity file and it is prepared according to the prescribed format of `SPCAT program'. This file 
contains maximum and minimum number of rotational states, partition function, 
rotational temperature and dipole moment of the molecule.
 
\subsection{Chemical Modeling}
In order to study various forms of choloronium in an
interstellar medium (ISM), we have developed a chemical model which includes
gas phase as well as grain surface chemical network. Our gas phase chemical network consists of
a network of \cite{wood07} and deuterated network used in
\citet{das13b}. In addition, we include some deuterated reactions by following \citet{robe00, 
albe13}. We include certain new reactions for
formation and destruction of various forms of choloronium and its related species.
Our present gas phase chemical network consists of 6180 reactions and present
surface chemical network consists of 285 reactions. Except molecular hydrogen
and Helium, depletion of all gas phase 
neutral species onto the grain surface are considered with a sticking probability of unity.
The reason for ignoring this for H$_2$ and He is that 
according to \citet{leit85}, the sticking coefficient of H$_2 \sim 0$ and  
\citet{robe00} argued that Helium would not stick to the grain at all.
Following is the list of reactions which are considered in our network 
for the formation/destruction of deuterated isotopomers of chloronium ion. 
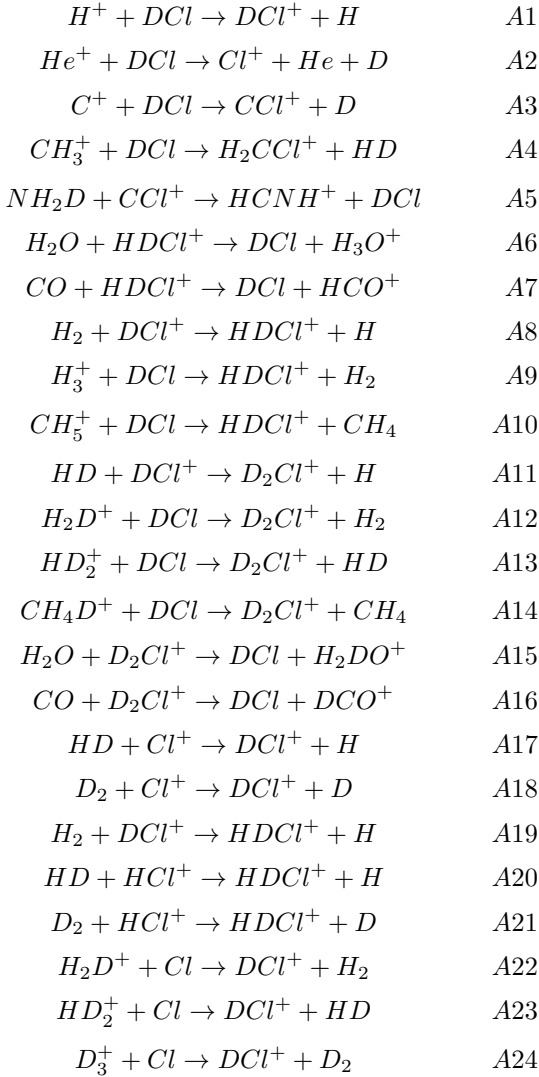
\begin{table*}
\centering{
\scriptsize
\caption{Initial abundances used relative to total hydrogen nuclei.}
\begin{tabular}{|c|c|}
\hline
Species&Abundance\\
\hline\hline
H$_2$ &    $5.00 \times 10^{-01}$\\
He    &    $1.00 \times 10^{-01}$\\
N     &    $2.14 \times 10^{-05}$\\
O     &    $1.76 \times 10^{-04}$\\
H$_3$$^+$&    $1.00 \times 10^{-11}$\\
C$^+$ &    $7.30 \times 10^{-05}$\\
S$^+$ &    $8.00 \times 10^{-08}$\\
Si$^+$&    $8.00 \times 10^{-09}$\\
Fe$^+$&    $3.00 \times 10^{-09}$\\
Na$^+$&    $2.00 \times 10^{-09}$\\
Mg$^+$&    $7.00 \times 10^{-09}$\\
P$^+$ &    $3.00 \times 10^{-09}$\\
Cl$^+$&    $4.00 \times 10^{-09}$\\
e$^-$ &    $7.31 \times 10^{-05}$\\
HD&  $1.6 \times 10^{-05}$\\
\hline
\end{tabular}}
\end{table*}
\begin{equation*}
H^+ + DCl\rightarrow DCl^+ + H
\eqno{A1}
\end{equation*}
\begin{equation*}
He^+ + DCl \rightarrow Cl^+ + He + D
\eqno{A2}
\end{equation*}
\begin{equation*}
C^+ + DCl \rightarrow CCl^+ + D
\eqno{A3}
\end{equation*}
\begin{equation*}
CH_3^+ + DCl \rightarrow H_2CCl^+ + HD
\eqno{A4}
\end{equation*}
\begin{equation*}
NH_2D + CCl^+ \rightarrow HCNH^+ + DCl
\eqno{A5}
\end{equation*}
\begin{equation*}
H_2O + HDCl^+ \rightarrow DCl + H_3O^+
\eqno{A6}
\end{equation*}
\begin{equation*}
CO + HDCl^+\rightarrow DCl + HCO^+
\eqno{A7}
\end{equation*}
\begin{equation*}
H_2 + DCl^+\rightarrow HDCl^+ + H
\eqno{A8}
\end{equation*}
\begin{equation*}
H_3^++ DCl\rightarrow HDCl^++ H_2
\eqno{A9}
\end{equation*}
\begin{equation*}
CH_5^++ DCl\rightarrow HDCl^++ CH_4
\eqno{A10}
\end{equation*}
\begin{equation*}
HD+ DCl^+\rightarrow D_2Cl^++ H
\eqno{A11}
\end{equation*}
\begin{equation*}
H_2D^++ DCl\rightarrow D_2Cl^++ H_2
\eqno{A12}
\end{equation*}
\begin{equation*}
HD_2^++ DCl\rightarrow D_2Cl^++ HD
\eqno{A13}
\end{equation*}
\begin{equation*}
CH_4D^++ DCl\rightarrow D_2Cl^++ CH_4
\eqno{A14}
\end{equation*}
\begin{equation*}
H_2O + D_2Cl^+\rightarrow DCl+ H_2DO^+
\eqno{A15}
\end{equation*}
\begin{equation*}
CO+ D_2Cl^+\rightarrow DCl+ DCO^+
\eqno{A16}
\end{equation*}
\begin{equation*}
HD+ Cl^+\rightarrow DCl^++ H
\eqno{A17}
\end{equation*}
\begin{equation*}
D_2+ Cl^+\rightarrow DCl^++ D
\eqno{A18}
\end{equation*}
\begin{equation*}
H_2+ DCl^+\rightarrow HDCl^++ H
\eqno{A19}
\end{equation*}
\begin{equation*}
HD+ HCl^+\rightarrow HDCl^++ H
\eqno{A20}
\end{equation*}
\begin{equation*}
D_2+ HCl^+\rightarrow HDCl^++ D
\eqno{A21}
\end{equation*}
\begin{equation*}
H_2D^++ Cl\rightarrow DCl^++ H_2
\eqno{A22}
\end{equation*}
\begin{equation*}
HD_2^++ Cl\rightarrow DCl^++ HD
\eqno{A23}
\end{equation*}
\begin{equation*}
D_3^++ Cl\rightarrow DCl^++ D_2
\eqno{A24}
\end{equation*}
 
Reaction A1 is a charge exchange type reaction. Rate coefficient for this reaction is assumed
to be similar to the rate coefficient adopted for $H^+ + HCl\rightarrow HCl^+ + H$ in
\citet{wood07}. Reactions A2 - A24 are Ion-neutral type reactions. 
According to \citet{herb06}, rate coefficients for ion-molecular reactions can be determined
by using capture theories (in which translational energy of reactants must only surpass a
long-range centrifugal barrier for a reaction to occur). The collision rate coefficient between an 
ion and non-polar neutral molecule can be determined by using so-called Langevin collision rate;
\begin{equation}
                 k = 2\pi e  \sqrt{\alpha_d/ \mu},
\end{equation}
where, $e$ is electronic charge, $\alpha_d$ is polarizabilty of neutral non-polar molecule,
$\mu$ is reduced mass of reactants. But for polar neutral species, a complex situation arises
due to attraction between a charge and a rotating permanent dipole moment.
In these cases, extensive trajectory calculations have been carried out by \citet{su82}
to predict rate coefficients ($k_{cap}$) of ion-polar molecule capture collisions.
According to \citet{woon09}, Su-Chesnavich formula can be written in two different ways
and both of which uses a parameter $x= \mu_D/\sqrt(2 \alpha k T)$ where k is Boltzman constant 
\& T is the temperature. The ion-dipole ($k_{cap}$) rates can be parameterized using 
following equations:
\begin{equation}
                    k_{cap}= (0.4767 x + 0.6200) k_L  ,
\end{equation}
and
\begin{equation}
               k_{cap}= [(x+0.5090)^2/10.526 + 0.9754] k_L.
\end{equation}
Equation 2 is used if $x \ge 2$ and Eqn. 3 is used if $x<2$.
Note that for $x=0$, it reduces to the Langevin expression.
Alternatively, the expression can be written in powers of temperature T. For example, when $x \geq 2$,
\begin{equation}
                            k_{cap}= c_1 + c_2 T^{-1/2},
\end{equation}
where,  $c_1= 0.62 k_L$ and $c_2= (0.4767 \mu_D/\sqrt(2 \alpha K) k_L$.
If the second term in Eqn. 4 is much greater than the first term, the expression
has $T^{-1/2}$ dependence which is used in both the UMIST and OSU databases.
We took values of polarizability and dipole moments of neutral species 
by following Woon \& Herbst (2009). They optimized equilibrium structures at
RCCSD(T)/aug-cc-pVTZ level of theory using finite field approach 
to obtain dipole moment and dipole polarizability components. According to their
calculations, values of the polarizabilities ($\alpha$) for HCl, H$_2$O, H$_2$, CO, NH$_3$ are to be 
2.538, 1.406, 0.773, 1.951 and 2.087 ${A\circ}^3$ respectively and values of the
Dipole moments ($\mu_D$) for HCl, H$_2$O, H$_2$, CO, NH$_3$ are to be 
1.075, 1.845, 0, 0.101, 1.519 Debye respectively. Now, computation of polarizability 
and dipole moments depend on the derivatives of electronic energy with respect to 
external electric field. This electronic energy is not dependent on the mass of the nuclei 
as calculations are based on Born-Oppenheimer approximation. 
But experimental difference arises mostly from difference in vibrationally averaged structure.
In our case, polarizability and dipole moments of HD, D$_2$, DCl, NH$_2$D are assumed to be similar
to their hydrogenated counter part. Plugging these values (polarizability and dipole moment of the
neutral species and reduced mass of the reactants) in above equations, we calculated 
reaction rates for the ion-neutral reactions (A2-A24).

Molecular ions could be destroyed by the following dissociative recombination reactions:
\begin{equation*}
DCl^++ e^-\rightarrow Cl+ D ,
\eqno{A25}
\end{equation*}
\begin{equation*}
D_2Cl^++ e^-\rightarrow Cl+ D+ D ,
\eqno{A26}
\end{equation*}
\begin{equation*}
D_2Cl^++ e^-\rightarrow DCl+ D,
\eqno{A27}
\end{equation*}
\begin{equation*}
HDCl^++ e^-\rightarrow Cl+ D+ H,
\eqno{A28}
\end{equation*}
\begin{equation*}
HDCl^++ e^-\rightarrow DCl+ H,
\eqno{A29}
\end{equation*}
In the absence of experimentally determined rate coefficients, 
following \citet{neuf09}, we use Eqns. 5 \& 6 for computation of DR rate 
coefficients (A25-A29) for diatomic and triatomic molecular ions, namely,
\begin{equation}
               k= 2\times 10^{-7} (T/300)^{-1/2}  cm^3 s^{-1},
\end{equation}
and
\begin{equation}
              k= 1.2\times 10^{-7} (T/300)^{-0.85} \ cm^3 s^{-1}.
\end{equation}

Following the destruction pathways of HCl \citep{wood07}, we have assumed that 
DCl could also be destroyed by Cosmic ray induced(CRPHOT) Photo reactions;
\begin{equation*}
DCl+ CRPHOT\rightarrow Cl+ D.
\eqno{A30}
\end{equation*}
The rate of Cosmic ray induced photo-reaction (A30) could be adopted as follows:
\begin{equation}
k_{CR}(T)= \alpha (T/300)^{\beta} \gamma/(1-\omega) \ s^{-1},
\end{equation}
where, cosmic-ray ionization rate is denoted by $\alpha$, photo reaction probability per cosmic ray 
ionization is denoted by $\gamma$ and $\omega$ denotes reflection coefficients of dust grain
in far UV. Following reaction $HCl + CRPHOT \rightarrow H + Cl$ in \citet{wood07}, 
here too we assume that $\alpha=1.3 \times 10^{-17}$, $\beta=0$, $\gamma=305$ and $\omega=0.6$.

DCl could also be dissociated by following interstellar photo reaction (PHOTON);
\begin{equation*}
DCl+ PHOTON\rightarrow Cl+ D .
\eqno{A31}
\end{equation*}
The rate coefficient for the reaction A31 could be calculated by using following relation:
\begin{equation}
k_{Phot}=\alpha exp(-\gamma A_V ) \ s^{-1},
\end{equation}
where, A$_V$ is the visual extinction and 
$\gamma=1.8$ is used following the reaction $HCl+PHOTON\rightarrow Cl+H$ in \citet{wood07}.

In our model, we assume that gas and grains are coupled through 
accretion and thermal/cosmic ray evaporation processes.
To model the environment of the outer edge of a dense interstellar cloud
or a diffuse cloud, we use T=10K, 
n$_H$ = $10^3-10^4$ cm$^{-3}$, A$_V=0.01-10$ magnitude. Following \citet{robe00}, 
in Table 1, initial elemental abundances relative to the total 
hydrogen nuclei is shown. This type of initial abundances are often adopted for the cold \& dark cloud.
All computed/estimated rate coefficients (A1-A31) are provided in Table 2 for T=10K, A$_V=10$.

\begin{table*}
\centering{
\scriptsize
\caption{Formation/destruction of deuterated chloronium ions and related species}
\begin{tabular}{|c|c|c|}
\hline
Reaction &Reaction Type& Rate coefficients \\
\hline\hline
$H^+ + DCl\rightarrow DCl^+ + H$ (A1)&Charge Exchange&$1.1\times 10^{-10}$ cm$^3s^{-1}$\\
$He^+ + DCl \rightarrow Cl^+ + He + D$ (A2)&Ion-neutral&$1.8\times 10^{-8}$ cm$^3s^{-1}$\\
$C^+ + DCl \rightarrow CCl^+ + D$ (A3)&Ion-neutral&$8.3\times 10^{-10}$ cm$^3s^{-1}$\\
$CH_3^+ + DCl \rightarrow H_2CCl^+ + HD$ (A4)&Ion-neutral&$7.7\times 10^{-09}$ cm$^3s^{-1}$\\
$NH_2D^+ CCl^+ \rightarrow HCNH^+ + DCl$ (A5)&Ion-neutral&$9.5\times 10^{-09}$ cm$^3s^{-1}$\\
$H_2O + HDCl^+ \rightarrow DCl + H_3O^+$ (A6)&Ion-neutral&$1.2\times 10^{-08}$ cm$^3s^{-1}$\\
$CO + HDCl^+\rightarrow DCl^+ HCO^+$ (A7)&Ion-neutral&$2.5\times 10^{-10}$ cm$^3s^{-1}$\\
$H_2 + DCl^+\rightarrow HDCl^+ + H$ (A8)&Ion-neutral&$1.5\times 10^{-09}$ cm$^3s^{-1}$\\
$H_3^++ DCl\rightarrow HDCl^++ H_2$ (A9)&Ion-neutral&$1.5\times 10^{-08}$ cm$^3s^{-1}$\\
$CH_5^++ DCl\rightarrow HDCl^++ CH_4$ (A10)&Ion-neutral&$7.3\times 10^{-09}$ cm$^3s^{-1}$\\
$HD+ DCl^+\rightarrow D_2Cl^++ H$ (A11)&Ion-neutral&$1.2\times 10^{-09}$ cm$^3s^{-1}$\\
$H_2D^++ DCl\rightarrow D_2Cl^++ H_2$ (A12)&Ion-neutral&$1.3\times 10^{-08}$ cm$^3s^{-1}$\\
$HD_2^++ DCl\rightarrow D_2Cl^++ HD$ (A13)&Ion-neutral&$1.2\times 10^{-08}$ cm$^3s^{-1}$\\
$CH_4D^++ DCl\rightarrow D_2Cl^++ CH_4$ (A14)&Ion-neutral&$7.2\times 10^{-09}$ cm$^3s^{-1}$\\
$H_2O + D_2Cl^+\rightarrow DCl+ H_2DO^+$ (A15)&Ion-neutral&$1.2\times 10^{-08}$ cm$^3s^{-1}$\\
$CO+ D_2Cl^+\rightarrow DCl+ DCO^+$ (A16)&Ion-neutral&$2.5\times 10^{-10}$ cm$^3s^{-1}$\\
$HD+ Cl^+\rightarrow DCl^++ H$ (A17)&Ion-neutral&$1.2\times 10^{-09}$ cm$^3s^{-1}$\\
$D_2+ Cl^+\rightarrow DCl^++ D$ (A18)&Ion-neutral&$1.1\times 10^{-09}$ cm$^3s^{-1}$\\
$H_2+ DCl^+\rightarrow HDCl^++ H$ (A19)&Ion-neutral&$1.5\times 10^{-09}$ cm$^3s^{-1}$\\
$HD+ HCl^+\rightarrow HDCl^++ H$ (A20)&Ion-neutral&$1.2\times 10^{-09}$ cm$^3s^{-1}$\\
$D_2+ HCl^+\rightarrow HDCl^++ D$ (A21)&Ion-neutral&$1.1\times 10^{-09}$ cm$^3s^{-1}$\\
$H_2D^++ Cl\rightarrow DCl^++ H_2$ (A22)&Ion-neutral&$1.0\times 10^{-09}$ cm$^3s^{-1}$\\
$HD_2^++ Cl\rightarrow DCl^++ HD$ (A23)&Ion-neutral&$1.0\times 10^{-09}$ cm$^3s^{-1}$\\
$D_3^++ Cl\rightarrow DCl^++ D_2$ (A24)&Ion-neutral&$1.0\times 10^{-09}$ cm$^3s^{-1}$\\
$DCl^++ e^-\rightarrow Cl+ D$ (A25)&Dissociative Recombination&$6.0\times 10^{-6}$ $s^{-1}$\\
$D_2Cl^++ e^-\rightarrow Cl+ D+ D$ (A26)&Dissociative Recombination&$1.2\times 10^{-5}$ $s^{-1}$\\
$D_2Cl^++ e^-\rightarrow DCl+ D$ (A27)&Dissociative Recombination&$1.2\times 10^{-5}$ $s^{-1}$\\
$HDCl^++ e^-\rightarrow Cl+ D+ H$ (A28)&Dissociative Recombination&$1.2\times 10^{-5}$ $s^{-1}$\\
$HDCl^++ e^-\rightarrow DCl+ H$ (A29)&Dissociative Recombination&$1.2\times 10^{-5}$ $s^{-1}$\\
$DCl+ CRPHOT\rightarrow Cl+ D$ (A30)&Cosmic ray induced Photodissociation&$9.9\times 10^{-15}$ $s^{-1}$\\
$DCl+ PHOTON\rightarrow Cl+ D$ (A31)&Photo-dissociation&$1.7\times 10^{-17}$ $s^{-1}$\\
\hline
\end{tabular}}
\end{table*}

\section{Results and Discussions}

\subsection{Chemical properties}
Interstellar chloronium (H$_2$Cl$^+$) is an asymmetric top with C$_{2V}$ symmetry. 
We performed quantum chemical simulation (MP2/aug-cc-pVQZ level of theory) for geometry 
optimization and energy calculation of chloronium ion. Ground state energies of 
H$_2$Cl$^+$ in gas and ice phases are found to be 
$-460.55382$ a.u, $-460.67002$ a.u. (1 atomic unit = $27.21$ eV) respectively. 
Due to solute-solvent electrostatic interaction (dipole level), ground state energy in 
ice phase is found to be lower than ground state energy in gas phase.
The solvent effect also brings changes 
in geometrical parameters of these species. Our results confirm that polarization of solute 
by continuum has important effects on absolute and relative solvation energies, which in turn 
changes the energy of respective species. The ground state energy of D$_2$Cl$^{+}$ and HDCl$^{+}$ are found 
to be similar to the ground state energy of H$_2$Cl$^{+}$ in gas and in ice phases. 
This is due to the fact that ground state energy and structure calculations are made by 
using Born-Oppenheimer approximation so isotopic mass is not a factor in the Hamiltonian. 
The dipole moment of chloronium in gas and in ice 
phases are found to be 1.98 Debye and 2.2024 Debye respectively.
The dipole moment of chloronium has been estimated as 1.89 Debye in {\it ab initio}
calculations performed by \citet{mull08}. According to \citet{puzz10}, additional splitting in
rotational spectrum is governed by nuclear quadrupole coupling of molecules. Moreover,
according to them, quantum chemical calculations could be extremely helpful for fine structure
analysis because it could provide electric-field gradient at corresponding nuclei.
In Table 3, we show a comparison of experiment and theoretical ground state quadrupole 
coupling constants of chlorine H$_2$Cl$^{+}$ for Mp2/aug-cc-pVQZ level of theory. Also, 
for the first time, we provide quadrupole coupling constants of chlorine in 
D$_2$Cl$^{+}$, HDCl$^{+}$ in the same table.

\begin{table*}
\centering{
\scriptsize
\caption{Theoretical and Experimental quadrupole coupling constants of chlorine in H$_2$Cl$^{+}$, 
D$_2$Cl$^{+}$, HDCl$^{+}$.}
\begin{tabular}{|c|c|c|c|}
\hline
Species&Constants&Theoretical values& Experimental values\\
&&{ in MHz}&{ in MHz$^a$} \\
\hline\hline
& $\chi_{aa}$ & -51.8851   & -53.44 \\
{\bf H$_2$Cl$^+$}& $\chi_{bb}$ & -14.3463  & -15.71    \\
& $\chi_{cc}$ & 66.2314  & 69.15    \\
& $\chi_{ab}$ & -0.4689  & -    \\
\hline
& $\chi_{aa}$ & -39.2014   & - \\
{\bf HDCl$^+$}& $\chi_{bb}$ & -27.1076  & -    \\
& $\chi_{cc}$ & 66.3090  & -    \\
& $\chi_{ab}$ & 17.4213  & -   \\
\hline
& $\chi_{aa}$ & -51.8863   & - \\
{\bf D$_2$Cl$^+$}& $\chi_{bb}$ & - 14.3470 & -    \\
& $\chi_{cc}$ & 66.2333  & -  \\
& $\chi_{ab}$ & -0.3328  & -    \\
\hline
\multicolumn{4}{|c|}{$^a$ \citet{sait88}}\\
\hline
\end{tabular}}
\end{table*}

\subsection{Chemical evolution and deuterium enrichment}
\begin{figure}
\vskip 1cm
\centering
\includegraphics[height=6cm,width=6cm]{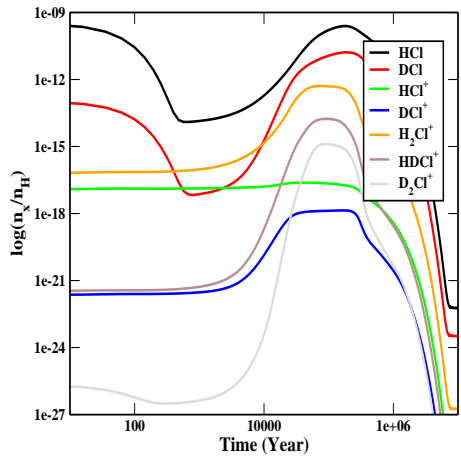}
\caption{Time evolution of the Chloronium ion and its related species.}
\label{fig-1}
\end{figure}

In Fig. 1, chemical evolution of chloronium ion and related species are shown. To mimic the
interstellar scenario, we consider n$_H=10^4$ cm$^{-3}$, A$_V=10$ and T=10K. 
It is assumed that initially all deuteriums are locked in the form of HD. 
Initial abundances of HD and H$_2$ are taken to be $1.6 \times 10^{-5}$ and $0.5$ respectively (Table 1) 
with respect to total number of hydrogen nuclei. This implies an initial fractionation ratio of
$3.2 \times 10^{-5}$. Unless otherwise stated, we use this initial fractionation ratio in all our simulations.
In our gas-grain model, we assume that all the neutral species could be depleted to icy
grain with a sticking coefficient of unity. Depleted species are allowed to populate the gas phase 
via thermal evaporation and cosmic ray evaporation processes. Due to depletion of  
neutral species, rate of production of their relative ions are decreased. Dissociative 
recombination processes then lead to destruction of 
these ionic species. As a result, abundances of these ions are decreased. 
In our simulation, we evolved our chemical code till a time comparable to 
the life time of a generic molecular cloud ($\sim$ $10^7$ years). From Fig. 1, it is clear 
that most of the species attain peak values near $\sim 10^5$ years. 
In the literature, a number of theoretical models attempted to find time variation of
physical properties of protostars since the beginning of
gravitational collapse \citep{smit98,myer98}. Time sequence of evolutionary stages
determined by various models are more or less similar but their estimates of absolute ages vary 
significantly \citep{empr08}. According to \citet{froe05}, 
the absolute age for Class 0/I borderline objects, for example, varies between 10$^4$ and a few times 10$^5$ 
years. Our simulation results in Fig. 1 shows that within this time frame, some species will attain peak 
values. It is also evident that one of the isotopomers of H$_2$Cl$^+$ (i.e., HDCl$^+$ from Fig. 1) 
is reasonably abundant within this time frame of collapse. So it is expected to be observed with ALMA, for example.

\begin{figure}
\centering
\includegraphics[height=8cm,width=8cm]{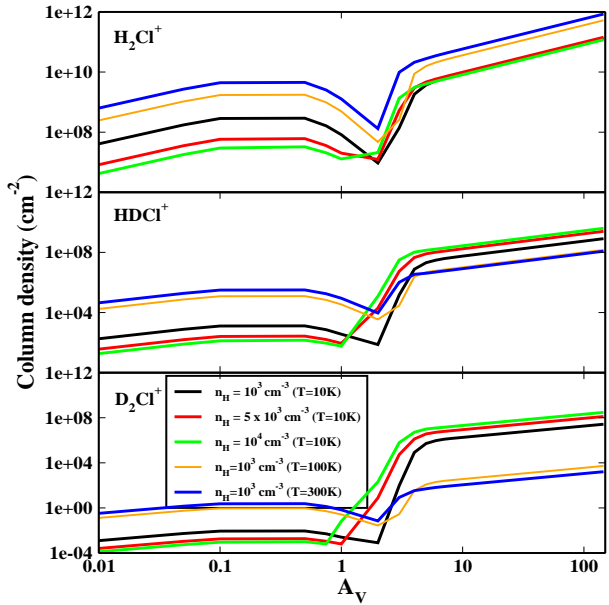}
\caption{Variation of peak column densities with visual extinction parameter (A$_V$). }
\label{fig-2}
\end{figure}

Following \citet{shal94} and \citet{das11}, the column density of a species could be calculated using the following relation:
\begin{equation}
N(A)=n_H x_i R,
\end{equation}
where, n$_H$ is the total hydrogen number density, x$_i$ is the abundance of the i$^{th}$
species and $R$ is the path length along the line of sight (=$\frac{1.6 \times 10^{21} \times A_V}{n_H}$).
According to \citet{lis10}, H$_2$Cl$^+$ is predicted to be the most abundant in those environments
where ultraviolet radiation is strong (i.e., in diffuse clouds, or near surfaces of dense
clouds that are illuminated by nearby O and B stars). In Fig. 2, we mimic these features
by varying visual extinction parameter (A$_V$) from $0.1$ to $145$. This high value of visual
extinction ($A_V=145$) could be used for W33A \citep{alla92}. Variation of peak column densities of
chloronium ion and its isotopomers are shown for different number density clouds (10$^3$-10$^4$ cm$^{-3}$). 
Abundance of H$_2$Cl$^+$ roughly remains constant beyond $A_V=5$. From Eqn. 9 it is clear that the
column density is directly proportional to $A_V$. As the abundance ($x_i$) remains constant beyond
$A_V=5$, peak column densities increase linearly. For A$_V < 4$, peak column densities are heavily affected. 
Column densities are maximum around heavily shielded region ($A_V>4$). 
\citet{lis10} computed column density of H$_2$Cl$^+$ to be $2.2-3.4 \times 10^{13}$ cm$^{-2}$
along NGC 6334I by assuming $5K$ excitation temperature.
\citet{neuf09} identified H$_2$Cl$^+$ around photo dissociation regions.
According to their prediction, column density of H$_2$Cl$^+$ around a photo 
dissociation region ($n_H \sim 10^4$ cm$^{-3}$, $\kappa_{UV} \sim 10^4$) would be $\sim 2.6 \times 10^{11}$ cm$^{-2}$.
$\kappa_{UV}$ is the UV radiation field which is normalized with respect to the
mean interstellar value by \citet{drai78}. 
\citet{neuf09} also predicted column density of choloronium $\sim 3 \times 10^{11}$ cm$^{-2}$ for
the conditions which could be applicable to the environment of Orion Bar PDR (n$_H \sim 6 \times 10^4$
cm$^{-3}$ and $\kappa_{UV}\sim 3 \times 10^4$). Their predicted column density was also 
$100$ times smaller than what is found in recent observation by \citet{lis10}. 
\citet{neuf12} mentioned that this theoretically
predicted value should not be compared directly with the observed values because theoretical
prediction was for the perpendicular column density whereas Orion Bar PDR was observed nearly edge-on.
They expect that geometrical enhancement in the observed value of the
column density by a factor of 4 could be possible.
In our case, we find a peak column density of H$_2$Cl$^+ \sim 1.3 \times 10^{11}$ cm$^{-2}$
in between $A_V=0.1 - 145$,  $n_H =10^3 -10^4$ cm$^{-3}$ and $T=10K$.
We also carried out our simulation at higher temperature to find out effects of
temperature on the formation of choloronium ion. For this case, we assumed
n$_H=10^3$ cm$^{-3}$ and varied visual extinction ($A_V$) in between $0.1 - 145$ 
for temperature 100K and 300K respectively (Fig. 2).
From Fig. 2, it is evident that column density of choloronium ion 
increases with temperature and we have the maximum column density of
$8.49 \times 10^{11}$ cm$^{-2}$ at 300K. Column densities of deuterated forms of choloronium ion 
are also heavily affected by increase in temperature. For $A_V<2$, column densities of these 
deuterated ions increases whereas for $A_V>2$ they are found to decrease when compared 
to the T=10K case. 

\begin{figure}
\centering
\includegraphics[height=6cm,width=6cm]{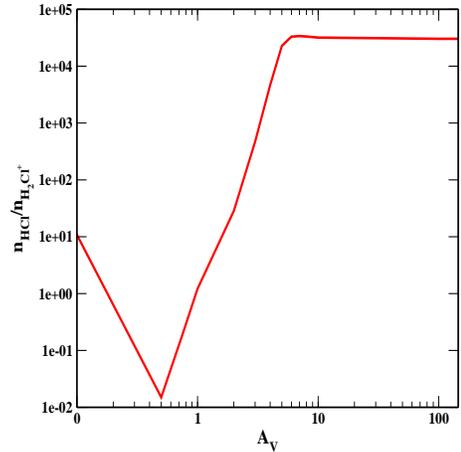}
\caption{Variation of HCl/H$_2$Cl$^+$ column density ratio with A$_V$}
\label{fig-3}
\end{figure}

\citet{lis10} derived HCl/H$_2$Cl$^+$ column density ratios to be in the range $\sim 1-10$, 
for diffuse and dense photon dominated regions (PDRs). In Fig. 3, we show variation of
HCl/H$_2$Cl$^+$ column density ratio with A$_V$. For A$_V<2$ (i.e., for the strong radiation 
field), our results are in line with observation. For this case, we assumed $n_H=10^4 \ cm^{-3}$,
$T=10$K, and HCl/H$_2$Cl$^+$ column density ratio was taken after the simulation time scale of $\sim 10^7$ year.

In Fig. 4, we show variation of fractionation ratio of HCl, HCl$^+$ and H$_2$Cl$^+$ with
initial $n_{HD/H_2}$ ratio. For this case, we assumed $n_H=10^4 \ cm^{-3}$, $A_V=10$ and $T=10$K.
HCl and HCl$^+$ are found to be heavily fractionated, always crossing elemental D/H 
ratio ($\sim 10^{-5}$, Linsky et al., 1995). It is interesting to note that in our
simulations, HDCl$^+$/H$_2$Cl$^+$ column density ratio is always above the elemental D/H ratio . 
Fractionation ratio of D$_2$Cl$^+$ molecule crossing the elemental D/H ratio beyond initial
$n_{HD/H_2}$ ratio $10^{-6}$. As H$_2$Cl$^+$ molecule is heavily fractionated,
we strongly suggest to look for HDCl$^+$ molecules in the same molecular 
clouds where H$_2$Cl$^+$ molecule was already been observed.

\begin{figure}
\vskip 3cm
\centering
\includegraphics[height=6cm,width=6cm]{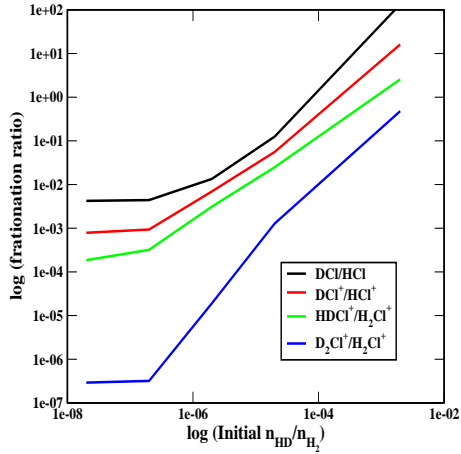}
\caption{Variation of fractionation ratio with initial  HD/H$_2$ ratio}
\label{fig-4}
\end{figure}

\subsection{Spectral Analysis}

We now turn to spectral properties of H$_2$Cl$^+$ and two of its isotopomers. For vibrational spectra, 
we compute infrared peak positions with their absorbance in gas phase as well as in water ice and mixed ice phases.
In Table 4, we present these for H$_2$Cl$^+$ and two of its isotopomers, namely, D$_2$Cl$^+$ and HDCl$^+$.
We find that the most intense mode of H$_2$Cl$^+$ in the gas phase
appears nearly at 2779.83 cm$^{-1}$. This peak is shifted in the ice phase
(water ice) by nearly 42 cm$^{-1}$ and appears at 2822.06 cm$^{-1}$.
The second strongest peak in the gas phase appears at 2789.60 cm$^{-1}$. It is also shifted in the ice phase.
To have a more realistic condition, instead of only the water ice,
we have considered a mixed ice mantle, which contains 70\% water, 20\% methanol and 10\% CO$_2$ molecules
\citep{kean01,das11,maju13a,maju13b,das13b}.
Gaussian 09W program uses a dielectric constant of water to be $\sim 78.5$.
For the case of mixed ice, we put the dielectric constant of the medium to be $61$, which is 
calculated by taking an weighted average of the dielectric
constants of H$_2$O, CH$_3$OH and CO$_2$. We note that
the most intense peak in gas phase is shifted in mixed ice also (Table 4).
Isotope effects on chemical shifts is caused by differences in
vibrational modes due to different isotope masses.
Each of the H$_2$Cl$^+$, D$_2$Cl$^+$ and HDCl$^+$ molecules has a unique spectrum because substitution of
isotope changes reduced mass of corresponding molecule.
We find that the most intense mode of D$_2$Cl$^+$ in gas phase appears
at 1995.63 cm$^{-1}$. This peak is shifted in ice
phase by 29.82 cm$^{-1}$, i.e., at 2025.45 cm$^{-1}$.
The second strongest peak in gas phase which appears at 1998.19 cm$^{-1}$
is also shifted in ice phase and appears at 2030.55 cm$^{-1}$.
The most intense peak in gas phase is similarly shifted in mixed solvated grain.
Infrared peak positions with their absorbance in gas phase as well as
in ice and mixed ice phases are marked for D$_2$Cl$^+$ and HDCl$^+$ in Table 4.
In Fig. 5 and in Fig. 6, we show how the isotopic substitution (D$_2$Cl$^+$ and HDCl$^+$) 
plays a part in vibrational progressions of H$_2$Cl$^+$ in gas phase and in ice phase.
Our results clearly show differences between spectroscopic parameters 
computed for interstellar chloronium ion in gas phase and in ice phase. 
These differences can be explained due to electrostatic 
effects that are often much less important for species placed in a solvent with high dielectric 
constant than they are in gas phase \citep{fore96}. 

\begin{figure}
\vskip 1cm
\centering
\includegraphics[height=6cm,width=6cm]{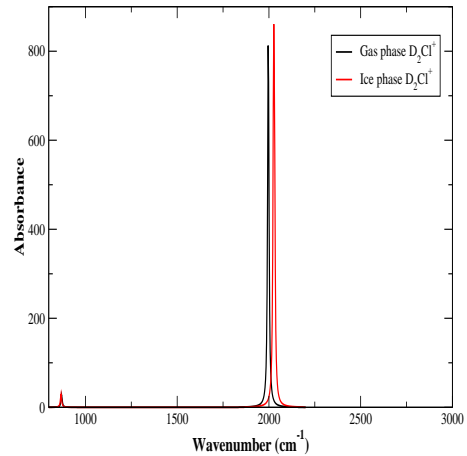}
\caption{Infrared spectrum of D$_2$Cl$^+$ in gas as well as in ice phase.}
\label{fig-5}
\end{figure}
\begin{figure}
\vskip 1cm
\centering
\includegraphics[height=6cm,width=6cm]{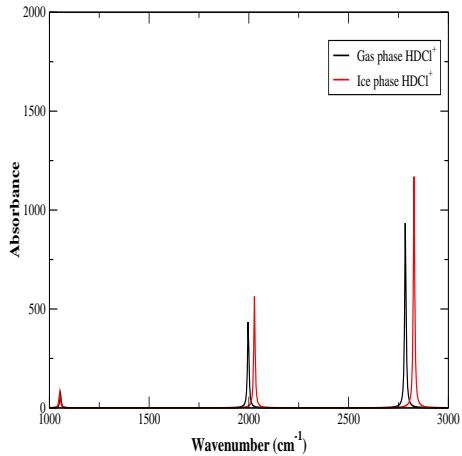}
\caption{Infrared spectrum of HDCl$^+$ in gas as well as in ice phase.}
\label{fig-6}
\end{figure}
In Table 5, we summarize our calculated theoretical values of rotational and quartic
centrifugal distortion constants for H$_2$Cl$^+$, D$_2$Cl$^+$ and HDCl$^+$ in gas phase. 
Calculated constants are corrected for each vibrational state as well as
vibrationally averaged structures. MP2/aug-cc-pVQZ level of theory has been used to perform 
these calculations for gas phase H$_2$Cl$^+$, D$_2$Cl$^+$ and HDCl$^+$. 
We compare our calculated spectroscopic constants (A, B, C, D$_{JK}$) of H$_2$Cl$^+$ with 
\citet{sait88} in Table 5. In order to summarize the outcome of our calculations about
rotational spectroscopy, we prepare our spectral information for one of the isotopologues of 
H$_2$Cl$^+$ (HDCl$^+$) as per guidelines of JPL (Table A1 of Appendix A). Computed transitions
are falling in between the 3 to 9 bands of ALMA (84 - 720 GHz). Our computed column
density also suggests that it could also be observed around the similar region, where H$_2$Cl$^+$ was observed.  

Different electronic absorption spectral parameters of H$_2$Cl$^+$ in gas phase are given in Table 6. 
In gas phase, the spectrum is characterized by five intense peaks at wavelengths of
$90.6$, $80.3$, $76.1$, $62.3$, $36.4$ nm (Fig. 7). These intense peaks are assigned
due to HOMO-LUMO transitions. These transitions correspond to
H-0$\rightarrow$ L+2, H-2$\rightarrow$ L+0, H-1$\rightarrow$ L+3, H-2$\rightarrow$ L+2, H-0$\rightarrow$ L+13.
Depending on the composition of interstellar grain mantle,
peak positions in ice phase are shifted. These features are presented 
in Table 6 with corresponding transition details. Fig. 7 clearly shows
some differences in gas phase and ice phase electronic absorption spectra of H$_2$Cl$^+$.

\begin{table*}
\scriptsize{
\centering
\vbox{
\caption{Vibrational frequencies of different forms of Chloronium in gas phase, in water
ice and in mixed ice at MP2/aug-cc-pVQZ level of theory}
\begin{tabular}{|c|c|c|c|c|c|c|c|c|}
\hline
\hline
{\bf Species}&{\bf Charge }&{\bf Spin}&{\bf Peak }&{\bf Absorbance}&{\bf Peak }&{\bf Absorbance}&{\bf Peak }&{\bf Absorbance}\\
{}&{ }&{\bf state}&{\bf positions }&{}&{\bf positions }&{}&{\bf positions}&{}\\
{}&{}&{}&{\bf (Gas phase)}&{}&{\bf (H$_2$O ice)}&{}&{\bf (Mixed ice)}&{}\\
&&&\bf (Wavenumber)&&\bf (Wavenumber)&& \bf (Wavenumber)& \\
&&&\bf ( in cm$^{-1}$)&&\bf ( in cm$^{-1}$)&& \bf ( in cm$^{-1}$)& \\
\hline
&&&1213.01&23.5567& 1208.84 & 25.3066 & 1209.44 & 24.9379 \\
{\bf H$_2$Cl$^+$}&{\bf Cation}&{\bf Singlet}& 2779.83& 342.2380& 2822.06 & 409.2761  & 2818.33 & 403.9254  \\
&&& 2789.60& 213.5581& 2834.00  & 267.9801  & 2831.03  & 262.6758 \\
\hline
&&& 870.12 & 8.0653 & 866.89 & 8.8043 & 867.34 & 8.6495 \\
{\bf D$_2$Cl$^+$}&{\bf Cation} & {\bf Singlet} & 1995.63 & 164.4252 & 2025.45 & 198.2665 & 2022.82 & 195.5651 \\
&&& 1998.19 & 100.4653 & 2030.55 & 127.2815 & 2028.39 & 124.6488\\
\hline
&&& 1055.58 & 21.0134 & 1051.30 & 23.6298 & 1051.20 &  23.2953 \\
{\bf HDCl$^+$}& {\bf Cation} & {\bf Singlet} & 1996.92 & 132.2521 & 2027.98 & 163.0641 & 2025.58 & 160.4419 \\
&&& 2784.69 & 278.3963 & 2828.05 & 338.3961 & 2824.70 & 333.0264\\
\hline
\hline
\end{tabular}}}
\end{table*}

\begin{table*}
\scriptsize{
\centering
\vbox{
\caption{Theoretical \& available Experimental rotational parameters of Chloronium and its different isotopomers at MP2/aug-cc-pVQZ level of theory
}
\begin{tabular}{|c|c|c|c|c|c|c|}
\hline
{\bf Species}&{\bf Rotational }&{\bf Values}&
{\bf Experimental}&
{\bf Distortional }&{\bf Values}&{\bf Experimental} \\
&{\bf constants }&{\bf in MHz}&
{\bf  values}&
{\bf  constants }&{\bf in MHz}&{\bf values} \\
&&&{\bf in MHz$^a$}&&&{\bf in MHz$^a$} \\
\hline
\hline
&A& 336730.77093& 337353.229 & $D_J$& 0.19899$\times$10$^{2}$&- \\
{\bf H$_2$Cl$^+$ in gas phase}& B &274941.86042& 273586.425 &$D_{JK}$& -0.73569$\times$10$^{2}$ & -71.814 \\
&C& 151335.22094& 148100.004 &$D_{K}$& 0.12577$\times$10$^{3}$&-    \\
& & && $d_1$& $-0.83773\times$10$^{1}$&- \\
&&&& $d_2$& 0.58014&-\\
\hline
&A& 309898.96477&   & $D_J$& 0.35636$\times$10$^{1}$& \\
{\bf HDCl$^+$ in gas phase}& B & 153561.70202 &  &$D_{JK}$& -0.92932$\times$10$^{1}$& \\
&C& 102697.60073& &$D_{K}$& 0.98535$\times$10$^{2}$&    \\
& & && $d_1$& $-0.13938\times$10$^{1}$& \\
&&&& $d_2$& -0.11650&\\
\hline
&A& 177672.18281 &  & $D_J$& 0.48691$\times$10$^{1}$& \\
{\bf D$_2$Cl$^+$ in gas phase}& B & 137566.23157&   &$D_{JK}$& -0.17934$\times$10$^{2}$& \\
&C& 77526.70474& &$D_{K}$& 0.33137$\times$10$^{2}$&    \\
& & && $d_1$& $-0.20906\times$10$^{1}$& \\
&&&& $d_2$& $-0.81717\times$10$^{-1}$ &\\
\hline
\multicolumn{7}{|c|}{$^a$ Saito \& Yamamoto et al., (1988)}\\
\hline
\end{tabular}}}
\end{table*}

\begin{table*}
\scriptsize{
\centering
\caption{Electronic transitions of H$_2$Cl$^+$ at B3LYP/6-311++G** level of theory}
\begin{tabular}{|c|c|c|c|c|c|}
\hline
\hline
{\bf Species}&{\bf Wavelength in nm}&{\bf Absorbance}&{\bf Oscillator strength }
&{\bf Transitions}&{\bf Contribution in \%}\\
\hline
&113.9&4864.07 &0.1201&H-1$\rightarrow$ L+0&98\\
& 99.7 & 5322.96&0.1316 &H-1$\rightarrow$ L+1&90\\
&90.6&11062.41&0.2728&H-0$\rightarrow$ L+2&100\\
& 80.3& 22614.49 &0.5051&H-2$\rightarrow$ L+0&66\\
&76.1&27125.91 &0.4172&H-1$\rightarrow$ L+3&70\\
& 70.8 & 9307.55&0.211 &H-2$\rightarrow$ L+1&57\\
{\bf H$_2$Cl$^+$ in gas phase}&62.3& 13154.01&0.2709&H-2$\rightarrow$ L+2&90\\
& 57.3& 1301.54 &0.03&H-0$\rightarrow$ L+9&100\\
& 53.3& 4644.87 &0.1117&H-1$\rightarrow$ L+8&96\\
& 44.2 & 4413.36&0.1047 &H-2$\rightarrow$ L+9&90\\
& 36.4&  9444.575 & 0.2395&H-0$\rightarrow$ L+13&100\\
& 34.5& 2347.95 &0.0594&H-1$\rightarrow$ L+13&98\\
&32.09& 1508.80 &0.0205&H-3$\rightarrow$ L+10&100\\
& 30.79 & 1549.00 &0.0402 &H-3$\rightarrow$ L+10&100\\
\hline
&  113.3  & 5297.52 &0.1308& H-1$\rightarrow$ L+0&98\\
& 98.1&11706.04&0.513&H-0$\rightarrow$ L+2&97\\
& 80.3&  31995.46&0.7694&H-2$\rightarrow$ L+0&66\\
&72.6&18217.29&0.4457&H-2$\rightarrow$ L+1&83\\
&  65.1 & 13505.80 &0 & H-2$\rightarrow$ L+4& 74\\
&  58.4&1618.54&0.0381&H-0$\rightarrow$ L+9&100\\
{\bf H$_2$Cl$^+$ in ice phase}& 55.0&  4755.99& 0.1215&H-1$\rightarrow$ L+8&96\\
& 46.8 & 2349.30& 0.0526&H-2$\rightarrow$ L+8&91\\
&  45.2  & 4262.98 & 0 & H-2$\rightarrow$ L+11&100\\
&  40.21&697.86  & 0.018 &H-2$\rightarrow$ L+12& 48\\
& 36.7&  9872.20 & 0.2521&H-0$\rightarrow$ L+13& 100\\
& 34.7 & 2272.75 & 0.065&H-1$\rightarrow$ L+13&99\\
&32.9& 1495.5532&0.0207&H-3$\rightarrow$L+11 & 100\\
& 31.4 & 1707.95 & 0.0437 & H-3$\rightarrow$L+12& 82\\ 
\hline 
&  113.5  & 5344.97 &0.1321& H-1$\rightarrow$ L+0&98\\
& 99.4& 206.40&0.1575&H-1$\rightarrow$ L+1&92\\
& 80.3&  31849.63&0.7566&H-2$\rightarrow$ L+0&63\\
{\bf H$_2$Cl$^+$ in mixed ice}&72.6&18544.85&0.4527&H-2$\rightarrow$ L+1&82\\
&  65.7 & 13565.30 &0 & H-2$\rightarrow$ L+5& 100\\
&  58.4&1616.83&0.0388&H-0$\rightarrow$ L+9&100\\
& 54.9&  4785.04& 0.1193&H-1$\rightarrow$ L+8&96\\
& 36.7 & 9833.04& 0.2511&H-0$\rightarrow$ L+13&100\\
\hline
\hline
\end{tabular}}
\end{table*}

\begin{figure}
\vskip 1cm
\centering
\includegraphics[height=6cm,width=6cm]{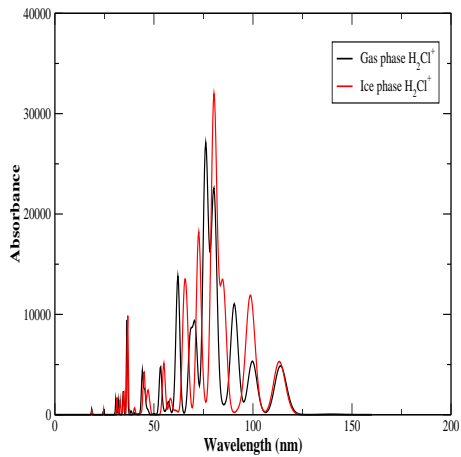}
\caption{Electronic absorption spectrum of H$_2$Cl$^+$ in gas phase and in ice phase.}
\label{fig-7}
\end{figure}

\section{Conclusions}

Protonated hydrogen chloride or chloronium (H$_2$Cl$^+$) ion has been recently identified 
towards Sgr A, W31C. In this paper, we investigated the existence and different aspects 
of its two isotopomers namely, HDCl$^+$ and D$_2$Cl$^+$. Following are the major results of this paper:

\noindent {$\bullet$ Langevin theory and parameterized trajectory theory are used for the computation
of various ion molecular reaction rates containing H$_2$Cl$^+$ and its related species.}

\noindent{$\bullet$ Our chemical modeling shows that HDCl$^+$ could be efficiently formed 
in gas phase and emission line should be strong enough to be observed.}

\noindent {$\bullet$ We explored vibrational, rotational and electronic 
spectral properties of H$_2$Cl$^+$, D$_2$Cl$^+$ and HDCl$^+$ in different astrophysical environments. 
Resulting chemical parameters could assist observers in identifying these molecules around Interstellar 
Molecular cloud. In Appendix A, we present a representative catalog files in JPL 
format for gas phase HDCl$^+$ which will be useful for observational purpose.}

\section{Acknowledgments}
Ankan Das is grateful to ISRO for financial support through a respond project 
(Grant No. ISRO/RES/2/372/11-12), SKC thank a DST project 
(Grant No.SR/S2/HEP-40/2008) and LM thank MOES for partial funding during this work.
The authors would like to thank the anonymous referee 
whose valuable suggestions have helped to improve this paper significantly.

\clearpage

{\hskip 8cm \bf Appendix-A}\\\\
{\noindent Existence of interstellar Chloronium ion has recently been reported by
Herschel Space Observatory$'$s Heterodyne Instrument for Far-Infrared \citep{neuf12}.
Our calculations show that one of the deuterated forms of choloronium ion (HDCl$^+$) is
significantly abundant and should be observed.
In order to summarize results of our computation on rotational spectroscopy,
we prepare our spectral information for HDCl$^+$ (Table A1 in the format of JPL catalog) 
to assist its detection around the ISM.
In Table A1, the computed rotational transitions for the gas phase HDCl$^+$ is shown. 
This Table is prepared with our calculated values of spectroscopic constants
which are given in Table 5. 
In Table A1, we have given the rotational transitions in MHz unit. Corresponding intensity
of lines are tabulated in nm$^2$ MHz unit. Intensity of any line represents
absorption cross section over spectral line shape (calculated at 300K and tabulated in base 10 
logarithm unit). Other parameters which are responsible for the computation of line frequencies are
given at footnote of Table A1.}

\begin{table*}
{\scriptsize
{\noindent {\bf Table A1:} Different rotational transitions and its related parameters
for gas phase HDCl$^+$ in the format of JPL catalog.}\\
\begin{tabular}{|c|c|c|c|c|c|c|c|c|c|}
\hline
{\bf Frequency$^a$} & {\bf Uncertainty$^b$ } & {\bf I$^c$} & {\bf D$^d$ } & {\bf E$_{lower}$$^e$} & {\bf g$_{up}$$^f$ } & {\bf Tag$^g$} & {\bf QnF$^h$} & {\bf Qn$_{up}$$^i$} & {\bf Qn$_{lower}$$^j$}\\
\hline
\hline
  257376.1993&  0.0000& -2.8248& 3 &  -0.0000&  4&  38001& 335& 1 0 1 2 2 &  0 0 0 2 1 \\
  257376.1993&  0.0000& -2.4316& 3 &  -0.0000&  4&  38001& 335& 1 0 1 2 2 &  0 0 0 2 2 \\  
  257376.1993&  0.0000& -2.8248& 3 &  -0.0000&  4&  38001& 335& 1 0 1 2 2 &  0 0 0 2 1 \\  
  257376.1993&  0.0000& -2.7830& 3 &  -0.0000&  4&  38001& 335& 1 0 1 2 2 &  0 0 0 2 3 \\       
  257376.2326&  0.0000& -2.7797& 3 &  -0.0000&  6&  38001& 335& 1 0 1 2 3 &  0 0 0 2 2 \\
  257376.2326&  0.0000& -2.0650& 3 &  -0.0000&  6&  38001& 335& 1 0 1 2 3 &  0 0 0 2 3 \\
  257376.2592&  0.0000& -2.7200& 3 &  -0.0000&  2&  38001& 335& 1 0 1 2 1 &  0 0 0 2 1 \\
  257376.2592&  0.0000& -2.8187& 3 &  -0.0000&  2&  38001& 335& 1 0 1 2 1 &  0 0 0 2 2 \\   
  257386.0031&  0.0000& -2.2865& 3 &  -0.0000&  4&  38001& 335& 1 0 1 3 2 &  0 0 0 2 1 \\
  257386.0031&  0.0000& -2.7946& 3 &  -0.0000&  4&  38001& 335& 1 0 1 3 2 &  0 0 0 2 2 \\  
  257386.0031&  0.0000& -4.1421& 3 &  -0.0000&  4&  38001& 335& 1 0 1 3 2 &  0 0 0 2 3 \\
  257386.0181&  0.0000& -1.8634& 3 &  -0.0000&  8&  38001& 335& 1 0 1 3 4 &  0 0 0 2 3\\
  257386.0529&  0.0000& -2.0650& 3 &  -0.0000&  6&  38001& 335& 1 0 1 3 3 &  0 0 0 2 2\\
  257386.0529&  0.0000& -2.7796& 3 &  -0.0000&  6&  38001& 335& 1 0 1 3 3 &  0 0 0 2 3 \\
  257393.8667&  0.0000& -2.8187& 3 &  -0.0000&  2&  38001& 335& 1 0 1 1 1 &  0 0 0 2 1\\
  257393.8667&  0.0000& -2.7199& 3 &  -0.0000&  2&  38001& 335& 1 0 1 1 1 &  0 0 0 2 2\\
  257393.8669&  0.0000& -3.7450& 3 &  -0.0000&  4&  38001& 335& 1 0 1 1 2 &  0 0 0 2 1\\
  257393.8669&  0.0000& -2.2901& 3 &  -0.0000&  4&  38001& 335& 1 0 1 1 2 &  0 0 0 2 3\\
  257393.8670&  0.0000& -2.8123& 3 &  -0.0000&  4&  38001& 335& 1 0 1 1 2 &  0 0 0 2 2\\
\hline
  514659.8572&  0.0000& -6.7938& 3 &   8.5857&  4&  38001& 335& 2 0 2 3 2 &  1 0 1 1 1\\
  514659.8824&  0.0000& -7.4094& 3 &   8.5857&  6&  38001& 335& 2 0 2 3 3 &  1 0 1 1 2\\
  514666.8692&  0.0000& -3.1365& 3 &   8.5857&  2&  38001& 335& 2 0 2 2 1 &  1 0 1 1 2\\
  514666.8694&  0.0000& -1.7143& 3 &   8.5857&  6&  38001& 335& 2 0 2 2 3 &  1 0 1 1 2\\
  514666.8695&  0.0000& -2.2435& 3 &   8.5857&  2&  38001& 335& 2 0 2 2 1 &  1 0 1 1 1\\
  514666.8696&  0.0000& -2.2437& 3 &   8.5857&  4&  38001& 335& 2 0 2 2 2 &  1 0 1 1 2\\
  514666.8698&  0.0000& -2.1448& 3 &   8.5857&  4&  38001& 335& 2 0 2 2 2 &  1 0 1 1 1\\
  514667.6710&  0.0000& -2.8243& 3 &   8.5855&  4&  38001& 335& 2 0 2 3 2 &  1 0 1 3 3\\
  514667.6788&  0.0000& -2.8208& 3 &   8.5855&  8&  38001& 335& 2 0 2 3 4 &  1 0 1 3 3\\
  514667.6964&  0.0000& -1.9589& 3 &   8.5855&  6&  38001& 335& 2 0 2 3 3 &  1 0 1 3 3\\
  514667.7136&  0.0000& -1.7698& 3 &   8.5855&  8&  38001& 335& 2 0 2 3 4 &  1 0 1 3 4\\
  514667.7208&  0.0000& -2.1126& 3 &   8.5855&  4&  38001& 335& 2 0 2 3 2 &  1 0 1 3 2\\
  514667.7313&  0.0000& -2.8220& 3 &   8.5855&  6&  38001& 335& 2 0 2 3 3 &  1 0 1 3 4\\
  514667.7463&  0.0000& -2.8260& 3 &   8.5855&  6&  38001& 335& 2 0 2 3 3 &  1 0 1 3 2\\
  514669.6441&  0.0000& -5.4590& 3 &   8.5857&  6&  38001& 335& 2 0 2 4 3 &  1 0 1 1 2\\
  514674.6834&  0.0000& -3.6631& 3 &   8.5855&  6&  38001& 335& 2 0 2 2 3 &  1 0 1 3 3\\
  514674.6836&  0.0000& -2.8877& 3 &   8.5855&  4&  38001& 335& 2 0 2 2 2 &  1 0 1 3 3\\
  514674.7183&  0.0000& -2.6833& 3 &   8.5855&  6&  38001& 335& 2 0 2 2 3 &  1 0 1 3 4\\
  514674.7331&  0.0000& -3.1158& 3 &   8.5855&  2&  38001& 335& 2 0 2 2 1 &  1 0 1 3 2\\
  514674.7332&  0.0000& -4.5605& 3 &   8.5855&  6&  38001& 335& 2 0 2 2 3 &  1 0 1 3 2\\
  514674.7334&  0.0000& -3.5845& 3 &   8.5855&  4&  38001& 335& 2 0 2 2 2 &  1 0 1 3 2\\
  514676.6697&  0.0000& -1.8438& 3 &   8.5857&  4&  38001& 335& 2 0 2 1 2 &  1 0 1 1 2\\
  514676.6698&  0.0000& -1.9425& 3 &   8.5857&  2&  38001& 335& 2 0 2 1 1 &  1 0 1 1 2\\
  514676.6699&  0.0000& -1.9425& 3 &   8.5857&  4&  38001& 335& 2 0 2 1 2 &  1 0 1 1 1\\
  514676.6700&  0.0000& -2.8364& 3 &   8.5857&  2&  38001& 335& 2 0 2 1 1 &  1 0 1 1 1\\
  514677.4582&  0.0000& -2.3000& 3 &   8.5855&  6&  38001& 335& 2 0 2 4 3 &  1 0 1 3 3\\
  514677.4648&  0.0000& -1.7918& 3 &   8.5851&  4&  38001& 335& 2 0 2 3 2 &  1 0 1 2 1\\
  514677.4913&  0.0000& -3.6483& 3 &   8.5851&  4&  38001& 335& 2 0 2 3 2 &  1 0 1 2 3\\
  514677.4930&  0.0000& -4.0096& 3 &   8.5855&  6&  38001& 335& 2 0 2 4 3 &  1 0 1 3 4\\
  514677.4991&  0.0000& -1.3636& 3 &   8.5851&  8&  38001& 335& 2 0 2 3 4 &  1 0 1 2 3\\
  514677.5064&  0.0000& -1.1122& 3 &   8.5855& 10&  38001& 335& 2 0 2 4 5 &  1 0 1 3 4\\
  514677.5080&  0.0000& -1.3848& 3 &   8.5855&  6&  38001& 335& 2 0 2 4 3 &  1 0 1 3 2\\
  514677.5167&  0.0000& -2.2922& 3 &   8.5851&  6&  38001& 335& 2 0 2 3 3 &  1 0 1 2 3\\
  514677.5172&  0.0000& -1.2463& 3 &   8.5855&  8&  38001& 335& 2 0 2 4 4 &  1 0 1 3 3\\
  514677.5246&  0.0000& -2.2754& 3 &   8.5851&  4&  38001& 335& 2 0 2 3 2 &  1 0 1 2 2\\
  514677.5500&  0.0000& -1.5647& 3 &   8.5851&  6&  38001& 335& 2 0 2 3 3 &  1 0 1 2 2\\
  514677.5521&  0.0000& -2.2942& 3 &   8.5855&  8&  38001& 335& 2 0 2 4 4 &  1 0 1 3 4\\
  514684.4770&  0.0000& -2.3423& 3 &   8.5851&  2&  38001& 335& 2 0 2 2 1 &  1 0 1 2 1\\
  514684.4773&  0.0000& -2.4262& 3 &   8.5851&  4&  38001& 335& 2 0 2 2 2 &  1 0 1 2 1\\
  514684.4837&  0.0000& -7.6805& 3 &   8.5855&  4&  38001& 335& 2 0 2 1 2 &  1 0 1 3 3\\
  514684.5037&  0.0000& -1.6815& 3 &   8.5851&  6&  38001& 335& 2 0 2 2 3 &  1 0 1 2 3\\
  514684.5039&  0.0000& -2.4062& 3 &   8.5851&  4&  38001& 335& 2 0 2 2 2 &  1 0 1 2 3\\
  514684.5335&  0.0000& -6.3471& 3 &   8.5855&  4&  38001& 335& 2 0 2 1 2 &  1 0 1 3 2\\
  514684.5337&  0.0000& -7.0346& 3 &   8.5855&  2&  38001& 335& 2 0 2 1 1 &  1 0 1 3 2\\
  514684.5368&  0.0000& -2.4322& 3 &   8.5851&  2&  38001& 335& 2 0 2 2 1 &  1 0 1 2 2\\
  514684.5370&  0.0000& -2.4096& 3 &   8.5851&  6&  38001& 335& 2 0 2 2 3 &  1 0 1 2 2\\
  514684.5372&  0.0000& -2.0557& 3 &   8.5851&  4&  38001& 335& 2 0 2 2 2 &  1 0 1 2 2\\
  514687.2785&  0.0000& -5.8721& 3 &   8.5851&  6&  38001& 335& 2 0 2 4 3 &  1 0 1 2 3\\
  514687.3117&  0.0000& -6.2690& 3 &   8.5851&  6&  38001& 335& 2 0 2 4 3 &  1 0 1 2 2\\
  514687.3375&  0.0000& -7.5999& 3 &   8.5851&  8&  38001& 335& 2 0 2 4 4 &  1 0 1 2 3\\
  514694.2775&  0.0000& -3.8539& 3 &   8.5851&  4&  38001& 335& 2 0 2 1 2 &  1 0 1 2 1\\
  514694.2776&  0.0000& -2.9352& 3 &   8.5851&  2&  38001& 335& 2 0 2 1 1 &  1 0 1 2 1\\
  514694.3040&  0.0000& -2.4123& 3 &   8.5851&  4&  38001& 335& 2 0 2 1 2 &  1 0 1 2 3\\
  514694.3373&  0.0000& -2.9434& 3 &   8.5851&  4&  38001& 335& 2 0 2 1 2 &  1 0 1 2 2\\
  514694.3374&  0.0000& -2.8496& 3 &   8.5851&  2&  38001& 335& 2 0 2 1 1 &  1 0 1 2 2\\
\hline
\end{tabular}}
\end{table*}
\clearpage
{\scriptsize
\begin{tabular}{|c|c|c|c|c|c|c|c|c|c|}
\hline
{\bf Frequency$^a$} & {\bf Uncertainty$^b$ } & {\bf I$^c$} & {\bf D$^d$ } & {\bf E$_{lower}$$^e$} & {\bf g$_{up}$$^f$ } & {\bf Tag$^g$} & {\bf QnF$^h$} & {\bf Qn$_{up}$$^i$} & {\bf Qn$_{lower}$$^j$}\\
\hline
\hline
  771769.9945&  0.0000& -7.6046& 3 &  25.7535&  6&  38001& 335& 3 0 3 4 3 &  2 0 2 1 2\\
  771774.5695&  0.0000& -6.9643& 3 &  25.7535&  4&  38001& 335& 3 0 3 3 2 &  2 0 2 1 1\\
  771774.5696&  0.0000& -6.7153& 3 &  25.7535&  4&  38001& 335& 3 0 3 3 2 &  2 0 2 1 2\\
  771776.9609&  0.0000& -2.9249& 3 &  25.7533&  6&  38001& 335& 3 0 3 4 3 &  2 0 2 4 4\\
  771776.9716&  0.0000& -2.9289& 3 &  25.7533& 10&  38001& 335& 3 0 3 4 5 &  2 0 2 4 4\\
  771777.0070&  0.0000& -1.7710& 3 &  25.7533&  8&  38001& 335& 3 0 3 4 4 &  2 0 2 4 4\\
  771777.0173&  0.0000& -1.6427& 3 &  25.7533& 10&  38001& 335& 3 0 3 4 5 &  2 0 2 4 5\\
  771777.0200&  0.0000& -1.8814& 3 &  25.7533&  6&  38001& 335& 3 0 3 4 3 &  2 0 2 4 3\\
  771777.0527&  0.0000& -2.9291& 3 &  25.7533&  8&  38001& 335& 3 0 3 4 4 &  2 0 2 4 5\\
  771777.0661&  0.0000& -2.9252& 3 &  25.7533&  8&  38001& 335& 3 0 3 4 4 &  2 0 2 4 3\\
  771779.7946&  0.0000& -8.0545& 3 &  25.7532&  6&  38001& 335& 3 0 3 4 3 &  2 0 2 2 2\\
  771779.7948&  0.0000& -5.5854& 3 &  25.7532&  6&  38001& 335& 3 0 3 4 3 &  2 0 2 2 3\\
  771779.8409&  0.0000& -6.5020& 3 &  25.7532&  8&  38001& 335& 3 0 3 4 4 &  2 0 2 2 3\\
  771781.5480&  0.0000& -4.2055& 3 &  25.7533&  8&  38001& 335& 3 0 3 3 4 &  2 0 2 4 4\\
  771781.5754&  0.0000& -3.0578& 3 &  25.7533&  6&  38001& 335& 3 0 3 3 3 &  2 0 2 4 4\\
  771781.5937&  0.0000& -2.9173& 3 &  25.7533&  8&  38001& 335& 3 0 3 3 4 &  2 0 2 4 5\\
  771781.5952&  0.0000& -3.2034& 3 &  25.7533&  4&  38001& 335& 3 0 3 3 2 &  2 0 2 4 3\\
  771781.6070&  0.0000& -7.8321& 3 &  25.7533&  8&  38001& 335& 3 0 3 3 4 &  2 0 2 4 3\\
  771781.6345&  0.0000& -3.9561& 3 &  25.7533&  6&  38001& 335& 3 0 3 3 3 &  2 0 2 4 3\\
  771784.3550&  0.0000& -1.6817& 3 &  25.7535&  2&  38001& 335& 3 0 3 2 1 &  2 0 2 1 1\\
  771784.3551&  0.0000& -2.5820& 3 &  25.7535&  2&  38001& 335& 3 0 3 2 1 &  2 0 2 1 2\\
  771784.3698&  0.0000& -1.9214& 3 &  25.7532&  4&  38001& 335& 3 0 3 3 2 &  2 0 2 2 2\\
  771784.3699&  0.0000& -3.2816& 3 &  25.7532&  4&  38001& 335& 3 0 3 3 2 &  2 0 2 2 3\\
  771784.3701&  0.0000& -1.4255& 3 &  25.7532&  4&  38001& 335& 3 0 3 3 2 &  2 0 2 2 1\\
  771784.3817&  0.0000& -1.1519& 3 &  25.7535&  6&  38001& 335& 3 0 3 2 3 &  2 0 2 1 2\\
  771784.3818&  0.0000& -0.9989& 3 &  25.7532&  8&  38001& 335& 3 0 3 3 4 &  2 0 2 2 3\\
  771784.4091&  0.0000& -1.1980& 3 &  25.7532&  6&  38001& 335& 3 0 3 3 3 &  2 0 2 2 2\\
  771784.4092&  0.0000& -1.9191& 3 &  25.7532&  6&  38001& 335& 3 0 3 3 3 &  2 0 2 2 3\\
  771784.4147&  0.0000& -1.5815& 3 &  25.7535&  4&  38001& 335& 3 0 3 2 2 &  2 0 2 1 1\\
  771784.4149&  0.0000& -1.6789& 3 &  25.7535&  4&  38001& 335& 3 0 3 2 2 &  2 0 2 1 2\\
  771786.7559&  0.0000& -2.0839& 3 &  25.7533&  8&  38001& 335& 3 0 3 5 4 &  2 0 2 4 4\\
  771786.7817&  0.0000& -2.0168& 3 &  25.7529&  6&  38001& 335& 3 0 3 4 3 &  2 0 2 3 3\\
  771786.7994&  0.0000& -3.6919& 3 &  25.7529&  6&  38001& 335& 3 0 3 4 3 &  2 0 2 3 4\\
  771786.8016&  0.0000& -3.9953& 3 &  25.7533&  8&  38001& 335& 3 0 3 5 4 &  2 0 2 4 5\\
  771786.8072&  0.0000& -1.1157& 3 &  25.7529&  6&  38001& 335& 3 0 3 4 3 &  2 0 2 3 2\\
  771786.8101&  0.0000& -0.8416& 3 &  25.7529& 10&  38001& 335& 3 0 3 4 5 &  2 0 2 3 4\\
  771786.8131&  0.0000& -0.6956& 3 &  25.7533& 12&  38001& 335& 3 0 3 5 6 &  2 0 2 4 5\\
  771786.8150&  0.0000& -0.8996& 3 &  25.7533&  8&  38001& 335& 3 0 3 5 4 &  2 0 2 4 3\\
  771786.8180&  0.0000& -0.7969& 3 &  25.7533& 10&  38001& 335& 3 0 3 5 5 &  2 0 2 4 4\\
  771786.8278&  0.0000& -0.9754& 3 &  25.7529&  8&  38001& 335& 3 0 3 4 4 &  2 0 2 3 3\\
  771786.8455&  0.0000& -2.0349& 3 &  25.7529&  8&  38001& 335& 3 0 3 4 4 &  2 0 2 3 4\\
  771786.8637&  0.0000& -2.0790& 3 &  25.7533& 10&  38001& 335& 3 0 3 5 5 &  2 0 2 4 5\\
\hline
\end{tabular}}
\clearpage
{\scriptsize
\begin{tabular}{|c|c|c|c|c|c|c|c|c|c|}
\hline
{\bf Frequency$^a$} & {\bf Uncertainty$^b$ } & {\bf I$^c$} & {\bf D$^d$ } & {\bf E$_{lower}$$^e$} & {\bf g$_{up}$$^f$ } & {\bf Tag$^g$} & {\bf QnF$^h$} & {\bf Qn$_{up}$$^i$} & {\bf Qn$_{lower}$$^j$}\\
\hline
\hline
  771789.5897&  0.0000& -5.6801& 3 &  25.7532&  8&  38001& 335& 3 0 3 5 4 &  2 0 2 2 3\\
  771791.3481&  0.0000& -7.9628& 3 &  25.7533&  6&  38001& 335& 3 0 3 2 3 &  2 0 2 4 4\\
  771791.3569&  0.0000& -2.5796& 3 &  25.7529&  4&  38001& 335& 3 0 3 3 2 &  2 0 2 3 3\\
  771791.3688&  0.0000& -2.5801& 3 &  25.7529&  8&  38001& 335& 3 0 3 3 4 &  2 0 2 3 3\\
  771791.3823&  0.0000& -1.8579& 3 &  25.7529&  4&  38001& 335& 3 0 3 3 2 &  2 0 2 3 2\\
  771791.3864&  0.0000& -1.5198& 3 &  25.7529&  8&  38001& 335& 3 0 3 3 4 &  2 0 2 3 4\\
  771791.3962&  0.0000& -1.7203& 3 &  25.7529&  6&  38001& 335& 3 0 3 3 3 &  2 0 2 3 3\\
  771791.4072&  0.0000& -6.1028& 3 &  25.7533&  6&  38001& 335& 3 0 3 2 3 &  2 0 2 4 3\\
  771791.4139&  0.0000& -2.5763& 3 &  25.7529&  6&  38001& 335& 3 0 3 3 3 &  2 0 2 3 4\\
  771791.4216&  0.0000& -2.5746& 3 &  25.7529&  6&  38001& 335& 3 0 3 3 3 &  2 0 2 3 2\\
  771791.4404&  0.0000& -6.5360& 3 &  25.7533&  4&  38001& 335& 3 0 3 2 2 &  2 0 2 4 3\\
  771794.1552&  0.0000& -2.3814& 3 &  25.7532&  2&  38001& 335& 3 0 3 2 1 &  2 0 2 2 2\\
  771794.1556&  0.0000& -2.2746& 3 &  25.7532&  2&  38001& 335& 3 0 3 2 1 &  2 0 2 2 1\\
  771794.1818&  0.0000& -2.3459& 3 &  25.7532&  6&  38001& 335& 3 0 3 2 3 &  2 0 2 2 2\\
  771794.1820&  0.0000& -1.6244& 3 &  25.7532&  6&  38001& 335& 3 0 3 2 3 &  2 0 2 2 3\\
  771794.2150&  0.0000& -1.9990& 3 &  25.7532&  4&  38001& 335& 3 0 3 2 2 &  2 0 2 2 2\\
  771794.2152&  0.0000& -2.3452& 3 &  25.7532&  4&  38001& 335& 3 0 3 2 2 &  2 0 2 2 3\\
  771794.2153&  0.0000& -2.3801& 3 &  25.7532&  4&  38001& 335& 3 0 3 2 2 &  2 0 2 2 1\\
  771796.5767&  0.0000& -6.7671& 3 &  25.7529&  8&  38001& 335& 3 0 3 5 4 &  2 0 2 3 3\\
  771796.5944&  0.0000& -6.3778& 3 &  25.7529&  8&  38001& 335& 3 0 3 5 4 &  2 0 2 3 4\\
  771796.6565&  0.0000& -7.5460& 3 &  25.7529& 10&  38001& 335& 3 0 3 5 5 &  2 0 2 3 4\\
  771801.1678&  0.0000& -3.1604& 3 &  25.7529&  2&  38001& 335& 3 0 3 2 1 &  2 0 2 3 2\\
  771801.1689&  0.0000& -3.6539& 3 &  25.7529&  6&  38001& 335& 3 0 3 2 3 &  2 0 2 3 3\\
  771801.1866&  0.0000& -2.7452& 3 &  25.7529&  6&  38001& 335& 3 0 3 2 3 &  2 0 2 3 4\\
  771801.1944&  0.0000& -5.1289& 3 &  25.7529&  6&  38001& 335& 3 0 3 2 3 &  2 0 2 3 2\\
  771801.2021&  0.0000& -2.9559& 3 &  25.7529&  4&  38001& 335& 3 0 3 2 2 &  2 0 2 3 3\\
  771801.2276&  0.0000& -3.6855& 3 &  25.7529&  4&  38001& 335& 3 0 3 2 2 &  2 0 2 3 2\\
\hline
\multicolumn{10}{|c|}{$^a$ Calculated frequency in MHz}\\
\multicolumn{10}{|c|}{$^b$ Calculated uncertainty of the line. If the line position is}\\ 
\multicolumn{10}{|c|}{in units of MHz then uncertainty of the line is greater or equal to zero.}\\
\multicolumn{10}{|c|}{$^c$ Base 10 logarithm of the integrated intensity at 300K in nm$^2$ MHz}\\
\multicolumn{10}{|c|}{$^d$ Degrees of freedom in the rotational partition function} \\
\multicolumn{10}{|c|}{(0 for atoms, 2 for linear molecules, 3 for non linear molecules)}\\
\multicolumn{10}{|c|}{$^e$ Lower state energy in cm$^{-1}$ relative to the lowest energy} \\
\multicolumn{10}{|c|}{level in the ground vibrionic state.}\\
\multicolumn{10}{|c|}{$^f$ Upper state degeneracy : g$_{up}=g_{I} \times g_{N}$, where g$_{I}$} \\
\multicolumn{10}{|c|}{is the spin statistical weight and g$_{N} =2N+1$ the rotational degeneracy.}\\
\multicolumn{10}{|c|}{$^g$ Molecule Tag}\\
\multicolumn{10}{|c|}{$^h$ Coding for the format of quantum numbers.}\\
\multicolumn{10}{|c|}{QnF=$100 \times Q + 10 \times H + N_{Qn}$; N$_{Qn}$ is the number of quantum }\\
\multicolumn{10}{|c|}{numbers for each state; H indicates the number of half integer quantum numbers;}\\ 
\multicolumn{10}{|c|}{Qmod5, the residual when Q is divided by 5, gives the number of principal}\\
\multicolumn{10}{|c|}{quantum numbers (without the spin designating ones)}\\
\multicolumn{10}{|c|}{$^i$ Quantum numbers for the upper state}\\
\multicolumn{10}{|c|}{$^j$ Quantum numbers for the lower state}\\
\hline
\end{tabular}}
%
\end{document}